\documentclass[letterpaper,10pt]{article} 
\usepackage{osameet3} %% use version 3 for proper copyright statement
\usepackage{multirow}
\usepackage{soulutf8}
\usepackage{booktabs} 
\usepackage{float}
\usepackage{aasmacros}
\usepackage{microtype}
\usepackage{bm}
\newcommand\authormark[1]{\textsuperscript{#1}}

\usepackage{amsmath,amssymb}
\usepackage[colorlinks=true,bookmarks=false,citecolor=blue,urlcolor=blue]{hyperref} %pdflatex

\begin{document}

\title{Oscillating Magnetized Color Superconducting Quark Stars}

\author{Marcos Osvaldo Celi,\authormark{1,2,*} Mauro Mariani,\authormark{1,2}  Milva Gabriela Orsaria\authormark{1,2}, and Lucas Tonetto \authormark{3,4}}

\address{\authormark{1} Grupo de Gravitaci\'on, Astrof\'isica y Cosmolog\'ia,
  Facultad de Ciencias Astron{\'o}micas y Geof{\'i}sicas, Universidad
  Nacional de La Plata, Paseo del Bosque S/N, La Plata 1900, Argentina\\
\authormark{2}CONICET, Godoy Cruz 2290, CABA 1425, Argentina\\
\authormark{3}Dipartimento di Fisica, ``Sapienza'' University of Rome, Piazzale A. Moro, 5. 00185 Roma, Italy \\
\authormark{4} INFN, Sezione di Roma, Piazzale A. Moro, 5. 00185 Roma, Italy}

\email{\authormark{*}mceli@fcaglp.unlp.edu.ar} %% email address is required?

\begin{abstract*}The main objective of this work is to study the structure, composition, and oscillation modes of color superconducting quark stars with intense magnetic fields. We adopted the MIT bag model within the color superconductivity CFL framework, and we included the effects of strong magnetic fields to construct the equation of state of stable quark matter. We calculated observable quantities, such as the mass, radius, frequency, and damping time of the oscillation fundamental $f$ mode of quark stars, taking into account current astrophysical constraints. The results obtained show that color superconducting magnetized quark stars satisfy the constraints imposed by the observations of massive pulsars and gravitational wave events. Furthermore, the quantities associated with the oscillation $f$ mode of these objects fit the universal relationships for compact objects. In the context of the new multi-messenger gravitational wave astronomy era and the future asteroseismology of neutron stars, we hope that our results contribute to the understanding of the behavior of dense matter and compact objects.
\end{abstract*}

\section{Introduction}\label{sec:introduction}

Strange matter, the possibility that the fundamental state of matter at high densities is a system of quarks in equilibrium against weak interactions, has been the subject of research in a wide variety of scenarios \citep{Chakrabarty:1989sqm, Saito:1990its, Madsen:1999hid, Weber:2005sqm, Felipe:2008msq, Han:2009sfs, Paulucci:2014sqm}. Probably one of the most favorable situations for the appearance of this type of matter, even without being absolutely stable, occurs during the last stages of the evolution of massive stars, associated with the explosion of type II supernovae, and during the formation of neutron stars (NSs) \citep{Dai:1995tco, Benvenuto:1999tpt, Sagert:2009sot, Bombaci:2011eoq, Malfatti:2019hqm}. The latter are the densest objects in the universe, with radii of just over a dozen kilometers and masses of around $1.4 M_{\odot}$. NSs, in addition to having high densities, have magnetic fields (MFs) from $10^{8}$ up to $10^{15}$~Gauss at their surface; in particular, the objects with the greatest MF values, $\sim 10^{14-15}$~Gauss, are the so-called \emph{magnetars}. Under these extreme conditions, QCD predicts that hadronic matter may undergo a transition to a strange matter phase or a \emph{color superconducting} phase \citep{Alford:2008csi}. This prediction, together with the \emph{stability of the strange matter hypothesis}, opens up the possibility of the formation of compact objects composed entirely of quark matter: \emph{quark stars} (QSs). The theoretical existence of such compact objects raises questions about what the true composition of NSs is, how astrophysical objects containing strange matter can be detected, and what are the main observational characteristics that differentiate them from objects made of non-strange matter. 

During the last decade, some NSs of $2 M_{\odot}$ were detected, establishing new restrictions to the \emph{equation of state} (EoS) of the matter that composes these objects and that is still unknown. The J1614-2230 pulsar is, among the most massive pulsars observed, the one with the least uncertainty in the determination of its mass, being $M = 1.906 \pm 0.016 M_{\odot}$~\mbox{\citep{Demorest:2010bx, Fonseca:2016tux, Arzoumanian:2018tny}}. Two other pulsars with a mass greater than two solar masses are J0348+0432, with a mass $M = 2.01 \pm 0.04 M_{\odot}$ \citep{Antoniadis:2013pzd}, and J0740+6620, with a mass $M = 2.08 \pm 0.07 M_{\odot}$ \citep{Fonseca:2021rfa}. Added to these data are those of the NICER telescope, from which not only the radius of the pulsar J0740+6620 was established, but also the mass and radius of the isolated pulsar J0030+0451 \citep{Riley:2019anv, Miller:2019pjm}.

On the other hand, the event called GW170817 corresponds to the detection of the first merger of two NSs, carried out by the LIGO/Virgo collaboration \citep{Abbott:2017oog}, which promoted the era of multi-messenger astronomy with \emph{gravitational waves} (GWs). This is one of the most promising scenarios to understand the behavior of the dense matter EoS. The analysis of the data from this event allowed for constraints on the mass, radius, and tidal deformability of the merging NSs. On 25 April 2019, a new merger of two NSs, GW190425, was detected, in which more massive NSs participated \citep{Abbott:2020goo}. The most recent event reported by the LIGO/Virgo collaboration, GW190814, was detected from a binary merger of a black hole and a $2.50$--$2.67 M_{\odot}$ compact object, which could be the most massive NS or the least massive black hole ever observed \citep{abbott2020gw190814}. It is expected that future GW detectors, such as the Einstein Telescope, could detect not only NSs mergers but the emission from the \emph{non-radial pulsation modes} of isolated NSs; in this context, it is also expected that the fundamental $f$ mode channels most of the GW emission energy \citep{Morozova:2018tgw}.

Due the uncertainty of the EoS of matter under extreme conditions of pressure and density and the constraints on NSs from multi-messenger astronomy, we propose that theoretical magnetized color superconducting QSs may shed some light on these questions. At this point, we must emphasize that we study magnetized color superconducting QSs under the assumption of the strange matter hypothesis. This is important because for the first time, a functional dependent magnetic field and the possibility of di-quark formation are both taken into account to study oscillating self-bound stars. Previous works used a uniform and constant MF to study magnetized strange quark matter within a confining model \citep{Sinha:2013sqm} and to study magnetized CFL stars in the framework of the MIT bag and NJL models \citep{2011IJMPE..20...84P, paulucci2011equation}. In particular, the NJL model does not satisfy the strange matter hypothesis \cite{BUBALLA_2005}. In fact, \citet{paulucci2011equation} relies on the MIT model analysis by adopting an effective bag by hand in the NJL to construct self-bound stars. Thinking that quark stars are possible, one of the problems with more realistic models to describe quark matter (see for example~\cite{G_mez_Dumm_2021} and the references therein) is the lack of free parameters to adjust them to astrophysical observations. In general, the parameters in this type of model are fixed, relying on vacuum fits, which might not be applicable at the large densities found in compact objects. An alternative model to study self-bound stars that has recently been revisited is the density-dependent quark mass model \cite{Backes_2021}. For simplicity, we adopted the traditional MIT bag model \citep{Chodos:1974nem} in our study. This allows us to fit the chosen free parameters not only to the latest astrophysical constraints for compact objects, but also to be consistent when modeling \emph{bare} self-bound QSs (without a surface crust layer \cite{Weber:2013soq}). Details of the construction of such objects and the results for the QSs' EoS, structure, and oscillation $f$ mode will be given in the next sections.

The paper is organized as follows. In Section~\ref{sec:mag}, we explain the incorporation of the MF within the MIT bag model. Section~\ref{sec:supercond} is devoted to a brief description of the color superconducting phase considered and to show the stability window of stable magnetized color superconducting matter, given the complete modeled EoS. Different families of compact objects, as well as the fundamental $f$ non-radial oscillation frequency and the corresponding damping times obtained are presented in Section~\ref{sec:tovf}. In Section~\ref{sec:conclus}, we present a summary and discussion of our work.

\section{Superconducting Magnetized Strange Quark Matter Equation of State}\label{sec:quarkeos}

\subsection{Magnetized Strange Quark Matter within the MIT Bag Model}\label{sec:mag}

In the presence of an MF, the transverse component of the momentum of a charged particle is quantized into \emph{Landau levels}. Assuming a local $z$-direction of the MF, this momentum reads \citep{Landau:1981qmn}:
\begin{equation}
 	k_\perp^2(\nu) = 2 \nu |q| B \, ,
\end{equation}
where $q$ is the electric charge of the particle, $B$ the strength of the MF, and $\nu$ a quantum number given by:
\begin{equation} 
\label{landaunu}
\nu = n + \frac{1}{2} - \text{sgn}(q) \frac{g}{2} \frac{s}{2} \, ,
\end{equation}
where $n$ is the Landau quantum number, $\text{sgn}(x)$ is the sign function, $g$ is the $g$-factor, and $s$ is the spin projection of the particle. We consider $g=2$ for spin $1/2$ particles.

The Landau quantization produces a energy spectrum that is
\begin{equation}
	E = \sqrt{k_z^2 + \bar{m}^2(\nu)} \, ,
\end{equation}
with
\begin{equation}
	\bar{m}^2(\nu) = m^2 + k_\perp(\nu)^2 \, .
\end{equation}
In order to keep $k_{z}$ real-valued,
\begin{equation}
	k_{z}(\nu) = \sqrt{E^2 - \bar{m}^2(\nu)} \, ,
\end{equation}
we have to impose the constraint $\nu \le \nu_{max}$ where,
\begin{equation}
	\nu_{\rm max} = \left\lfloor \frac{E^2 - m^2}{2|q|B} \right\rfloor ,
\label{numax} 
\end{equation}
and $\lfloor x \rfloor$ is the largest integer less than or equal to $x$. 

The anisotropy of the MF and, consequently, of the momentum components induces an energy--momentum tensor for the matter component given by:
\begin{equation}
 T_{\mu \nu}^{\textrm{matter}}= \textrm{diag}(\varepsilon, P_\perp,P_\perp,P_\parallel) \, ,
\end{equation}
where $\varepsilon$ is the energy density,
and the pressure components read
\begin{eqnarray}
 P_\parallel&=&-\Omega \, , \nonumber \\
 P_\perp&=& -\Omega- \mathcal{M} B \nonumber \, ,
\end{eqnarray}
$\mathcal{M}$ being the matter magnetization \citep{Blanford:1982mso,Felipe:2008msq}:
\begin{equation}
 \mathcal{M}= - \partial \Omega / \partial B \rvert_{\mu_B}\, .
\label{magmatter}
\end{equation}
Besides, the pure contribution from the MF generates another anisotropy in the system:
\begin{equation}
T_{\mu \nu}^{\textrm{MF}} = \textrm{diag}(B^2/2, B^2/2, B^2/2, -B^2/2) \, .
\end{equation}
In addition, in the \emph{MIT bag model}, there appears a constant contribution to the energy density and the pressure, which mimics the confinement property of QCD \citep{Chodos:1974nem,Chodos:1974bsi}. This contribution is treated as a free parameter of the model, $Bag$, and is given by
\begin{equation}
 T_{\mu \nu}^{Bag}= \textrm{diag}(Bag, -Bag,-Bag,-Bag) \, .
\end{equation}
Hence, the total energy--momentum tensor of the system turns out to be
\begin{equation}
T_{\mu \nu}= T_{\mu \nu}^{\textrm{matter}}+T_{\mu \nu}^{Bag}+T_{\mu \nu}^{\textrm{MF}} \, .
\end{equation}

As we already have stated in previous works \citep{Mariani:2019mhs, Mariani:2022omh} and following the work by \citet{strickland:2012bpo}, in a quark matter system under a locally constant MF in the $z$-direction, the integrals of thermodynamic quantities are substituted by sums over the transverse momentum due to the quantization. Thus, the particle number density, energy density, and pressures of each i-particle of the system are given by
\begin{eqnarray}
n^i&=&\frac{\gamma_{c}\left|q\right| B}{2 \pi^{2}} \sum_{-s}^{+s} \sum_{n=0}^{\nu\leq \nu_{\max }} k_{z, F} \, , \label{eq:termomag} \nonumber\\ 
\varepsilon^i&=&\frac{\gamma_{c}\left|q\right| B}{4 \pi^{2}} \sum_{-s}^{+s} \sum_{n=0}^{\nu\leq \nu_{\max }}\left[E_{\mathrm{F}} k_{z, F}+\bar{m}^{2} \ln \left(\frac{E_{\mathrm{F}}+k_{z, F}}{\bar{m}}\right)\right] \, , \nonumber \\
P^i_{\|}&=&\frac{\gamma_{c}\left|q\right| B}{4 \pi^{2}} \sum_{-s}^{+s} \sum_{n=0}^{\nu\leq \nu_{\max }}\left[E_{\mathrm{F}} k_{z, F}-\bar{m}^{2} \ln \left(\frac{E_{\mathrm{F}}+k_{z, F}}{\bar{m}}\right)\right] \, , \nonumber \\
P^i_{\perp}&=&\frac{\gamma_{c}\left|q\right|^{2} B^{2}}{2 \pi^{2}} \sum_{-s}^{+s} \sum_{n=0}^{\nu\leq \nu_{\max }} \nu\ln \left(\frac{E_{\mathrm{F}}+k_{z, F}}{\bar{m}}\right) \, ,
\label{Eqs:magnetic}
\end{eqnarray}
with $E_F=\mu_i$ ($\mu_i$ being the chemical potential of the particle) and $k_{z, F}$ being the Fermi energy and $z$-momentum, respectively. The factor $\gamma_{c}=3$ indicates the degeneracy color number of the~quarks.

The determination of the MF direction and strength inside NSs is a very complex problem, which implies the resolution of the non-linear general relativistic magneto-hydrodynamic equations \citep{Pili:2014aem}. Magneto-hydrodynamics models have shown that, during the proto-NS stage, the differential rotation would develop high toroidal MF components inside the star~\citep{Bonano:2003mfd,Naso:2008mfa,Frieben:2012emo}, so both the poloidal and toroidal MF components are necessary to preserve the star stability \citep{Ciolfi:2012pfi}. Thus, realistic stable models of magnetized NSs require the simultaneous presence of poloidal and toroidal MF components \citep{braithwaite2006evolution,Ciolfi:2013ttc,Sur:2020mfc}. Besides, purely toroidal MFs make the NS prolate, while purely poloidal MFs tend to make it oblate; if both toroidal and poloidal components are of the same order, we may expect that oblateness and prolateness cancel out approximately, leading to stars close to spherical symmetry. In this scenario, known as the \emph{chaotic MF} approximation \citep{zel2014stars,Flores:2020gws}, a spatial average can be performed and the spherical symmetry of the total system remains unchanged. Consequently, we can define an effective isotropic pressure given by \citep{Bednarek:2003tio,Flores:2016pos, Mariani:2019mhs}:
\begin{equation}
	P=\frac{T_{1 1}+T_{2 2}+T_{3 3}}{3}= \sum_{i=u,d,s} \frac{2 P^i_\perp+P^i_\parallel}{3} - Bag + \frac{B^2}{6} \, .
\label{pressure_prescription}	
\end{equation}

Furthermore, when aiming to study the structure and composition of magnetized compact objects, it is usual to avoid the complex MF dynamics and distribution using a functional form to model the MF strength profile in a given direction. \mbox{\citet{Dexheimer:2017wis}} used a polynomial MF profile in the star's polar direction, satisfying Maxwell's equations. A polynomial poloidal MF profile in magnetars has also been suggested by \mbox{\citet{Chatterjee:2019mfd}}. These MF parametrizations are constructed consistently since they satisfy the Maxwell equations, and so, they provide a suitable physics insight into the MF profile inside NSs. However, the accuracy of these profiles in a direction other than polar and their compatibility with the presence of a non-negligible toroidal component have not yet been studied. Furthermore, these profiles are almost flat, and considering a surface value of $B \sim 10^{15}$~Gauss---corresponding to the observed MF surface values for magnetars---they do not allow reaching internal MF strength values much beyond this order of magnitude. This feature prevents reaching MFs of $\sim 10^{17}$--$10^{18}$~Gauss in the center of magnetars, values that emerge from magneto-hydrodynamics simulations \citep{Igoshev:2021eon,Pili:2014aem, Frieben:2012emo,Naso:2008mfa}; in particular, some works even suggest that the central MF in hybrid stars with quark matter in their cores could be as large as $10^{19}$~Gauss \citep{Sotani:2015mhq} and in QSs could be as large as $10^{20}$~Gauss \citep{Ferrer:2010eos, Chu:2018qma, Chu:2021qsm}.

In this work, we adopted a hypothetical functional form of the MF parametrization depending on the baryonic chemical potential, $\mu_B$, given by 
\begin{equation}
 B(\mu_B) = B_{min} + B_{max} \left(1 - e^{(\beta(({\mu_B}-m_{n})^{\alpha})/m_{n})}\right) \, ,
\label{Eqs:B_profile}
\end{equation}
with $\alpha=2.5$ and $\beta = -4.08 \times 10^{-4}$ and $m_n$ is the nucleon mass \citep{Dexheimer:2012hsi}. The parameters $B_{\text{min}}$ and $B_{\text{max}}$ correspond to the order of magnitude of the MF at the surface and the center of the star, respectively. We define two sets for these parameters to analyze two paradigmatic NS scenarios: the \emph{low MF} pulsar and the \emph{magnetar} (see the details in Table~\ref{table:bcases}). As can be seen in that table, in order to study a wide range of MF values, we selected the value of $B_{max} \lesssim 3 \times 10^{18}$~Gauss according to the constraint imposed by \citet{Lai:1991ces}, who found that greater MF values could destabilize the star.

The exponential profile of Equation~(\ref{Eqs:B_profile}), although not physically consistent, is a suitable choice to study bulk properties such as the structure and composition of these compact objects since it allows covering a wide range of MF values. Although hypothetical, this exponential parametrization has been used in several works and is an acceptable approximation to model the complex and still unknown profile of the MF \citep{Bandyopadhyay:1997qmf,Mao:2003aso,Rabhi:2009qhp,Dexheimer:2012hsi,Flores:2020gws,Thapa:2020eos, Mariani:2022omh}.

\begin{table} [H]
	\caption{MF parametrization values for the two selected astrophysical scenarios: \emph{low-MF} pulsar and~\emph{magnetar}.}
	\label{table:bcases}
	\centering
\setlength{\tabcolsep}{13.6mm}
	\begin{tabular}{cccc} 
		\toprule
		Scenario & $B_{\text{min}}$~(Gauss) & $B_{\text{max}}$~(Gauss) \\
		\midrule
		Low-MF & $1 \times 10^{13}$ & $1 \times 10^{15}$ \\
		Magnetar & $1 \times 10^{15}$ & $3 \times 10^{18}$ \\
		\bottomrule
	\end{tabular}

\end{table}

\subsection{Color Superconductivity and Stability Window}\label{sec:supercond}

According to the QCD phase diagram, it is likely that, at high densities and low temperatures, strange matter becomes a color superconductor (see for example \citep{Rajagopal:1999mtq, Fukushima:2010tpd, Guenther:2021oot}). Any attractive quark--quark interaction in this type of regime would lead to the appearance of condensates called di-quarks, similar to electron condensates in ordinary superconductivity~\citep{Bardeen:1957mto}. Thus, conventionally, di-quarks are zero momentum spineless Cooper~pairs. 

In electromagnetic superconductivity, if the MF applied to the system is greater than a critical value, $B_c$, Cooper pairs could break and the system is reverted to a normally conducting state. This happens because the electrons in the Cooper pair have equal charges and opposite spins; thus, they have anti-parallel magnetic moments to the MF. The MF will tend to align the two parallel magnetic moments closest to each other, destroying the superconducting state. This transition from the superconducting to the normal state depends on whether the superconductor is of first or second order. In the first-order superconductor, the superconducting state has an abrupt transition to the normal state when $B > B_c$. This type of superconductor is characterized by completely expelling the magnetic field, which is known as the Meissner effect. Second-order superconductors have two critical fields, $B_{c1}$ and $B_{c2}$; for $B < B_{c1}$, the superconducting state holds, and for $B > B_{c2}$, there exists the normal state. Between $B_{c1}$ and $B_{c2}$, there is a mixed state in which the flux tubes of the MF may penetrate the superconductor \cite{glampedakis2011magnetohydrodynamics}. In the case of color superconductivity, quarks forming di-quarks have opposite charges and spins. Thus, the magnetic moments are parallel to the MF, and therefore, its presence reinforces the color superconductivity \cite{Ferrer_2006}.

In NSs, superconductivity may be accompanied by the baryon superfluidity and/or the electromagnetic Meissner effect \cite{Shovkovy_2005}, and the MF can be expelled during a long time period \cite{baym1969spin} or may exist in vortices \cite{Haskell_2018}. In the case of pairing in P-wave states, the superfluidity/superconductivity may remain for $B > B_{c2}$, as in ferromagnetic superconductors and in some color superconducting phases \cite{PhysRevD.101.056011}.

The analysis of the pairing properties for strange matter as a color superconductor is not trivial due to the variety of flavors, colors, and quark masses involved. In addition, the formation of a color superconducting phase breaks the SU(3)$_c$ color symmetry, so di-quarks are not colorless. Therefore, in the framework of QSs, it will not only be necessary to guarantee the electric charge neutrality, but also the color charge neutrality.

\begin{figure}[H]
\centering
\includegraphics[width=0.5\linewidth]{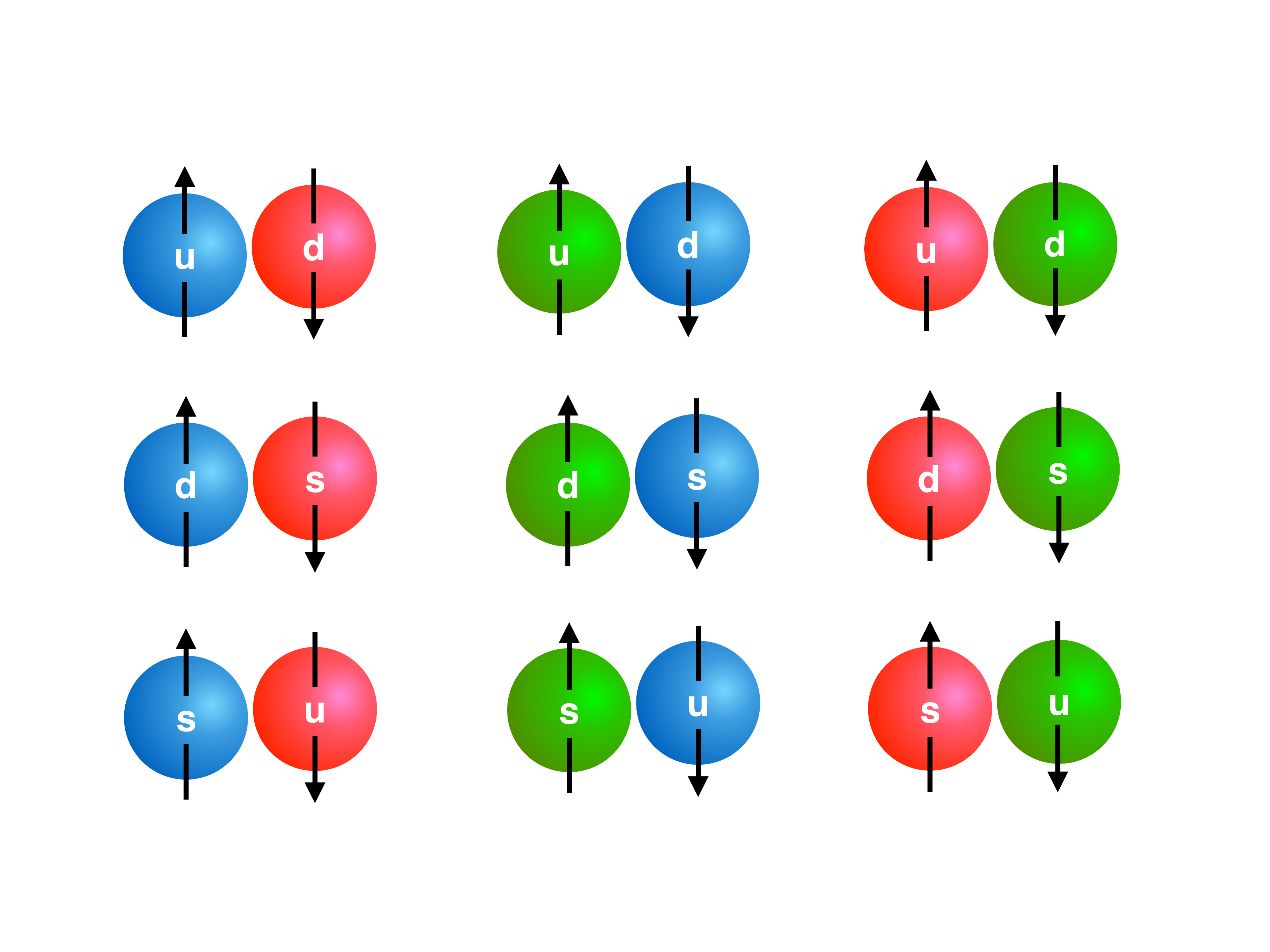}
\caption{Schematic illustration of the different pairing combinations in the CFL phase. Quarks forming Cooper pairs have opposite charges and spins, and the presence of an MF reinforces the pairing (see text).}
\label{fig:pattern}
\end{figure}

One of the most-studied color superconducting phases is~\emph{color flavor locked} (CFL)~\citep{Alford:2008csi}, and it has also been studied considering the presence of an MF \citep{Noronha:2007cfl,Fukushima:2008csm}. This state is a symmetric phase of matter in which all the light quarks, each with three colors, are involved in the pairing process. A schematic representation of the pairing patterns in this phase is shown in Figure~\ref{fig:pattern}. The formation of di-quarks lowers the energy of the system by an amount related to the so-called color superconducting gap, $\Delta$. This quantity is a function of the chemical potential, but can be treated as a free parameter of the model. To include the effect of color superconductivity in a phenomenological way, we considered a fictional condition of unpaired quark matter that is in a superconducting state once the quarks involved in the pairing reach a common Fermi momentum (see, for example, \citet{Curin:2021dse} and the references therein). The energy of the system is affected by the term
\begin{equation}
\varepsilon_{\Delta}=-3\left(\frac{\Delta \bar{\mu} }{ \pi}\right)^2, 
\end{equation}
where 
\begin{equation}
 \overline{\mu} = \frac{1}{N} \sum_{i} {\mu}_{i},
\end{equation}
is the mean chemical potential related to the $N$ quarks forming di-quarks. The quark chemical potentials considering electric and color charges are given by
\begin{equation}
 \mu_i=\mu_B-Q\mu_e+T_3\mu_3+T_8\mu_8 \, ,
\end{equation}
where
\begin{eqnarray}
 Q &=& \textrm{diag}(2/3,-1/3,-1/3) \, , \nonumber \\ 
 T_3 &=& \textrm{diag}(1/2,-1/2,0) \, , \nonumber \\ 
 T_8 &=& \textrm{diag}(1/3,1/3,-2/3) \, \nonumber.
\end{eqnarray}
$Q$ is the diagonal matrix corresponding to the electric charge. The color potentials $\mu_3$ and $\mu_8$ are associated with the two color charges of the group SU(3)$_c$ that commute with each other, and $T_3$ and $T8$ are the color charge diagonal matrices, the generators of SU(3)$_c$.

Using the Euler relation, the complete EoS of magnetized color superconducting quark matter is given by 
\begin{equation}
\varepsilon = - (P-\varepsilon_{\Delta}) + \sum_{i=u,d,s,e,\mu}\mu_i n^i,
\end{equation}
where $P$ is given by Equation~\eqref{pressure_prescription}, and the number densities $n^i$ of Equation~\eqref{Eqs:magnetic} are modified by the di-quark formation fulfilling the following relationship:
\begin{equation}
 n^u=n^d=n^s=n_B \, .
\label{Eq:densi_rel}
\end{equation}
where, now, $n^i = \partial(P-\varepsilon_{\Delta})/\partial \mu_i$ and $n_B$ is the baryonic number density of the system. Equation~\eqref{Eq:densi_rel} is a consequence of imposing both color and electric charge neutrality, the latter given by
\begin{equation}
 2n^u-n^d-n^s=3 (n^e+n^{\mu}) \, .
\label{neut_electrica}
\end{equation} 

It is important to mention that the CFL phase can undergo stresses due to the mass of the strange quark and the conditions of charge neutrality and beta equilibrium. Furthermore, for this superconducting phase to exist, the pairing between the quarks must be strong enough, {{i.e}}, $\Delta > m_{s}^2/2 \mu$, becoming unstable in the limit where $\Delta \sim m_{s}^2/2 \mu$. In particular, \emph{gapless} CFL (gCFL) replaces the CFL phase at $m_s^2/2 \mu > \Delta$ \citep{alford2005hybrid, Alford:2008csi, Alford_2005}. We obtained that for the two extreme values of the superconducting gap considered, $\Delta$ = 10, 90 MeV, the transition from CFL to gCFL should occur for chemical potentials lower than $\mu$ = 460.8, 51.2 MeV, respectively, well below the chemical potential values from which the EoSs of Table \ref{tabla:tablaedes} are physical, that is when the pressure increases monotonically with the energy density. In more realistic quark models, such as NJL, the gCFL phase may have astrophysical implications, controlling the cooling of a neutron star if quark matter in this phase is present~\cite{Alford_2005_cooling}.

\begin{figure}[H]
\centering
 \includegraphics[width=0.75\linewidth]{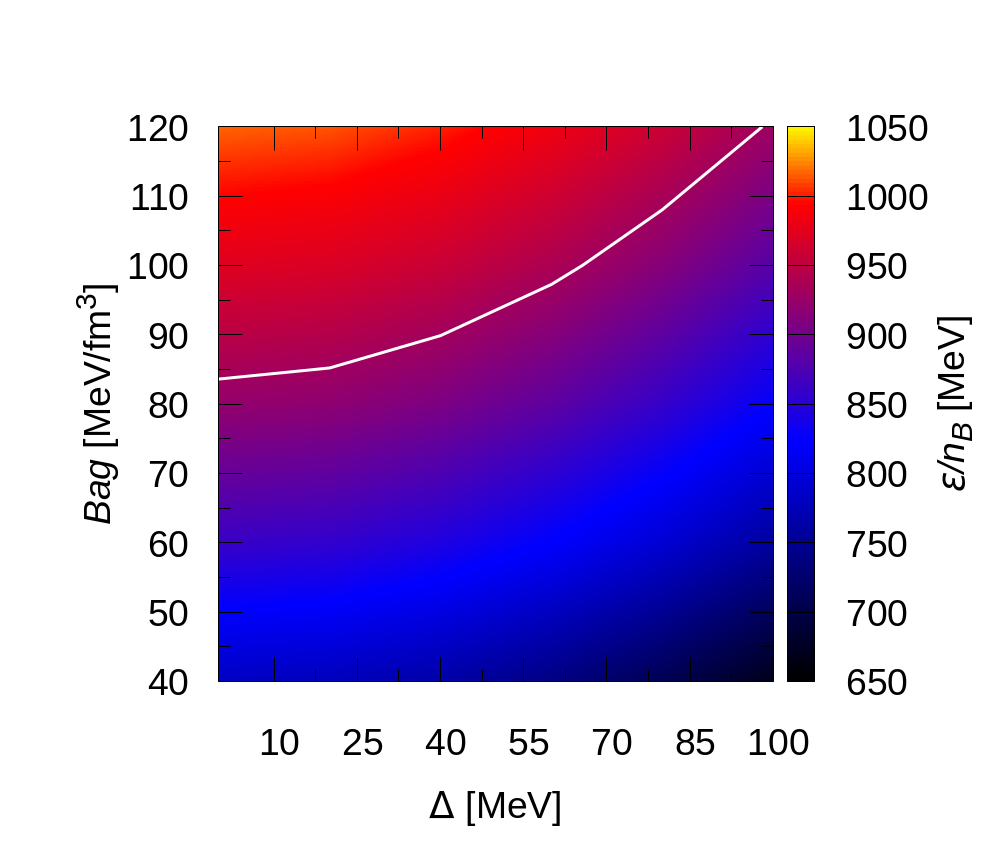}
 \caption{Energy per baryon, $\varepsilon/n_B$, as a color map in the $Bag$--$\Delta$ plane considering a low and constant MF, $B=10^{12}$~Gauss. The white curve indicates the constraint corresponding to the mass of $^{56}$Fe. Only $Bag$--$\Delta$ pairs below the white curves lead to stable configurations.}
 \label{fig:estab}
\end{figure}

For QSs to exist, strange quark matter composing them must fulfill the absolute stability hypothesis. Therefore, to obtain the corresponding stability window, we calculated the energy per baryon of superconducting magnetized quark matter, $\varepsilon/n_B$, at pressure $P=0$, which corresponds to the conditions on the QSs surface. Due to the observational constraints imposed for NSs with low MFs, we used a constant MF, $B=10^{12}$~Gauss, for the stability analysis. Hence, once we have built the superconducting magnetized quark EoS, firstly, we will study the stability of quark strange matter within our model. In Figure~\ref{fig:estab}, we present the energy density per baryon as a function of the free parameters of the model: the \textit{Bag} constant and the superconducting gap $\Delta$. Through this analysis, we can find the stability window of our model, which in the figure corresponds to the region below the white curve (the $^{56}$Fe mass). As can be seen, we obtain stable strange quark matter for any value of $\Delta$ as long as we keep the value of $Bag$ low enough.

\vspace{-12pt}
{}

\begin{table}[H]
\centering
%\resizebox{0.6\textwidth}{!}
\caption{Parameters for the four chosen EoSs. All the sets satisfy the strange matter stability hypothesis and the mass constraint $M_{\textrm{max}}\geq2.01 M_\odot$. We also show the surface baryonic density, $n_B^{sur}$, and the central MF strength, $B_{cen}$, for each set.}
\label{tabla:tablaedes}
{
\setlength{\tabcolsep}{3.4mm}
\begin{tabular}{ccccccc}
\toprule
EoS \# & \begin{tabular}[c]{@{}c@{}}$\Delta$\\ {(}MeV{)}\end{tabular} & \begin{tabular}[c]{@{}c@{}} $Bag$ \\ $($MeV/fm$^3)$\end{tabular} & \begin{tabular}[c]{@{}c@{}} $\varepsilon/n_B\rvert_{P=0}$ \\ {(}MeV{)}\end{tabular} & \begin{tabular}[c]{@{}c@{}} $M_{max}$ \\ $(M_\odot)$ \end{tabular} & \begin{tabular}[c]{@{}c@{}} $n_B^{sur}$ \\ $($1/fm$^3)$ \end{tabular} & \begin{tabular}[c]{@{}c@{}} $B_{cen}$ \\ $(10^{18}$~Gauss$)$ \end{tabular} \\ \midrule
1 & 10 & 45 & 801.9 & 2.17 & 0.23 & 1.4\\ 
2 & 50 & 50 & 794.8 & 2.18 & 0.24 & 1.4\\ 
3 & 90 & 45 & 712.0 & 2.60 & 0.21 & 0.6\\ 
4 & 90 & 70 & 809.9 & 2.03 & 0.30 & 1.7\\ \bottomrule
\end{tabular}
}

\end{table}
\newpage
\section{Solutions of the Structure Equations and $f$ Oscillation Mode}\label{sec:tovf}

To calculate the structure of color superconducting magnetized QSs, we assumed that these objects have spherical symmetry and do not rotate. Under this hypothesis, it is valid to integrate the well-known relativistic hydrostatic equilibrium equations of Tolman, Oppenheimer, and Volkov (TOV) and obtain several stellar properties for different equations of state. Furthermore, since isolated compact objects can oscillate, we calculated the oscillation frequencies and the associated damping times of their non-radial oscillation modes, because these types of modes emit GWs. We only focused on the fundamental \textit{f} mode, since it concentrates the largest amount of potentially detectable energy.

\begin{figure}[H]
\centering
 \includegraphics[width=0.75\linewidth]{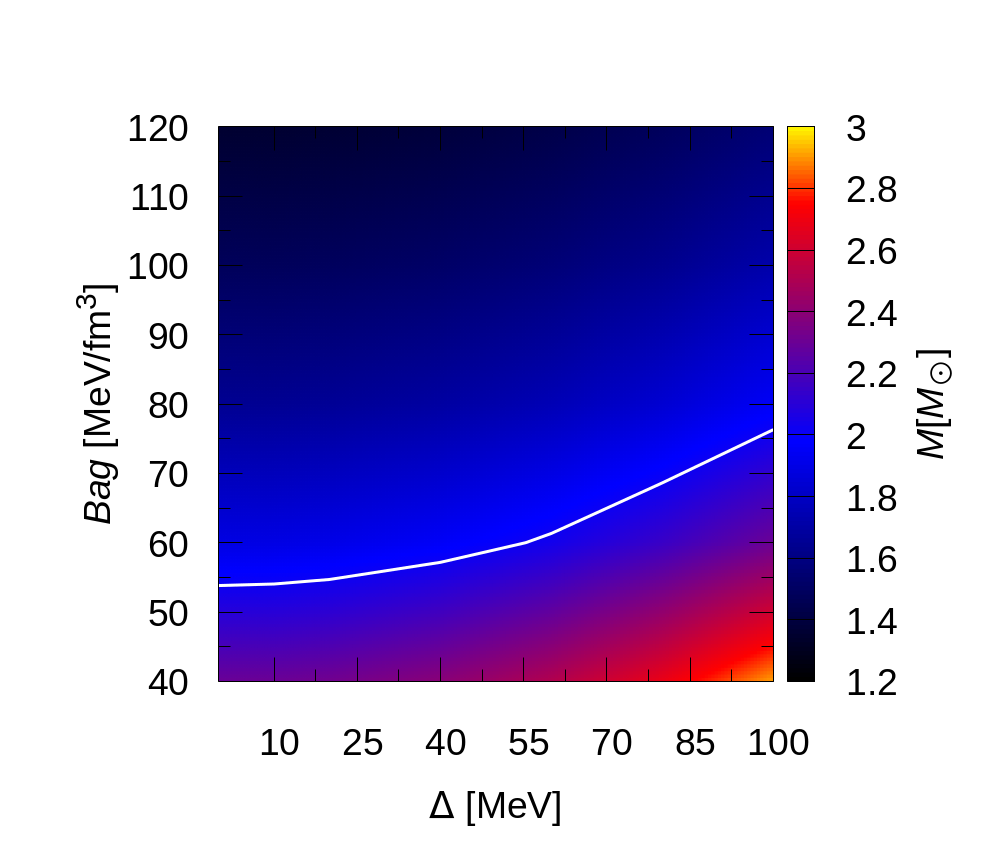}
 \caption{Maximum mass for each QS family, $M_{max}$, in the $Bag$--$\Delta$ plane considering the low MF parametrization. The white curve indicates the constraint of massive pulsars, $M_{max} = 2.01 M_\odot$.}
 \label{fig:masamax}
\end{figure}

In addition to satisfying the stability hypothesis, the QS families obtained with the superconducting magnetized quark EoSs must fulfill the $2.01 M_\odot$ constraint imposed by the observation of the pulsar J0740+6620 \citep{Arzoumanian:2018tny, Fonseca:2021rfa}. In Figure~\ref{fig:masamax}, we present the QS maximum mass (for a low-MF parametrization; see Table~\ref{table:bcases}) as a function of the $Bag$ and $\Delta$ parameters, and the white curve represents the $M_{max} = 2.01 M_\odot$ restriction. Any combination ($Bag$, $\Delta$) below this curve satisfies such a constraint.

%\vspace{-12pt}
{}

\begin{figure}[H]
\centering
 \includegraphics[width=0.6\linewidth]{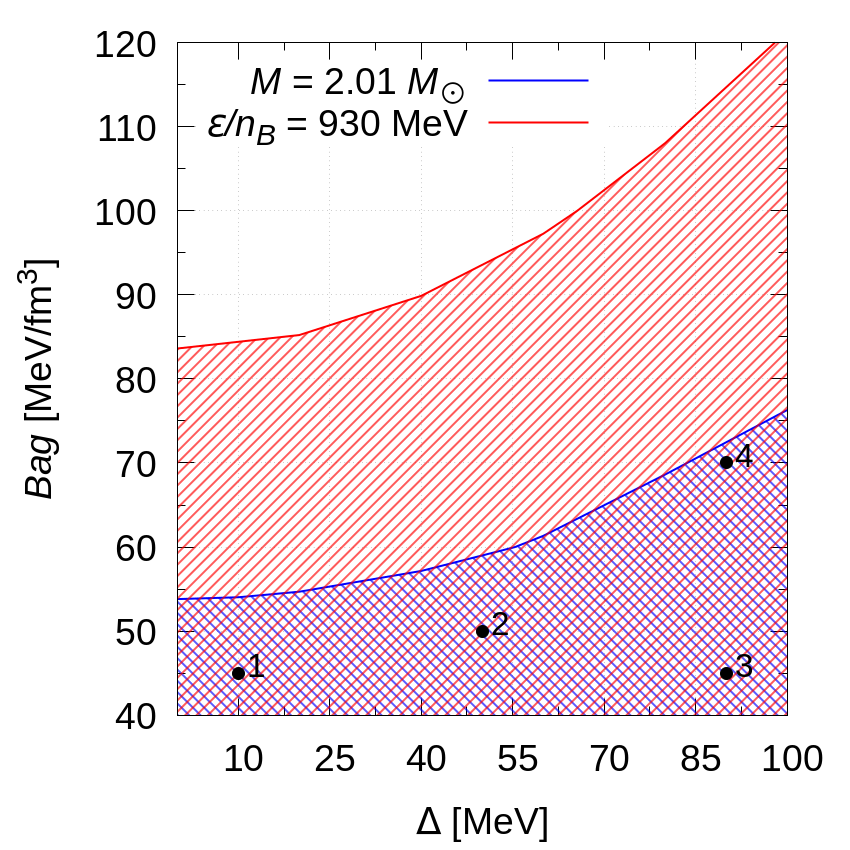}
 \caption{$Bag$--$\Delta$ plane showing the stability window and the $M_{max} \ge 2.01 M_\odot$ allowed regions. We selected four qualitatively representative sets (Table~\ref{tabla:tablaedes}) of the general behavior of our model from the overlapping region where both conditions are satisfied.}
 \label{fig:eleccioneos}
\end{figure}

In Figure~\ref{fig:eleccioneos}, we show the combination of both the stability window of magnetized superconducting quark matter and the constraint of massive pulsars, Figures~\ref{fig:estab} and \ref{fig:masamax}, respectively. We selected four sets of parameters, indicated with black numbered dots (see Table~\ref{tabla:tablaedes} for details), from the overlapped region. The choice of sets also allow us to study the effects of $Bag$ and $\Delta$ parameters in a decoupled way: Sets~$1$ and $3$ (Sets~$3$ and $4$) share the same value of $Bag$ ($\Delta$).

\begin{figure}[H]
\centering
 \includegraphics[width=0.6\linewidth]{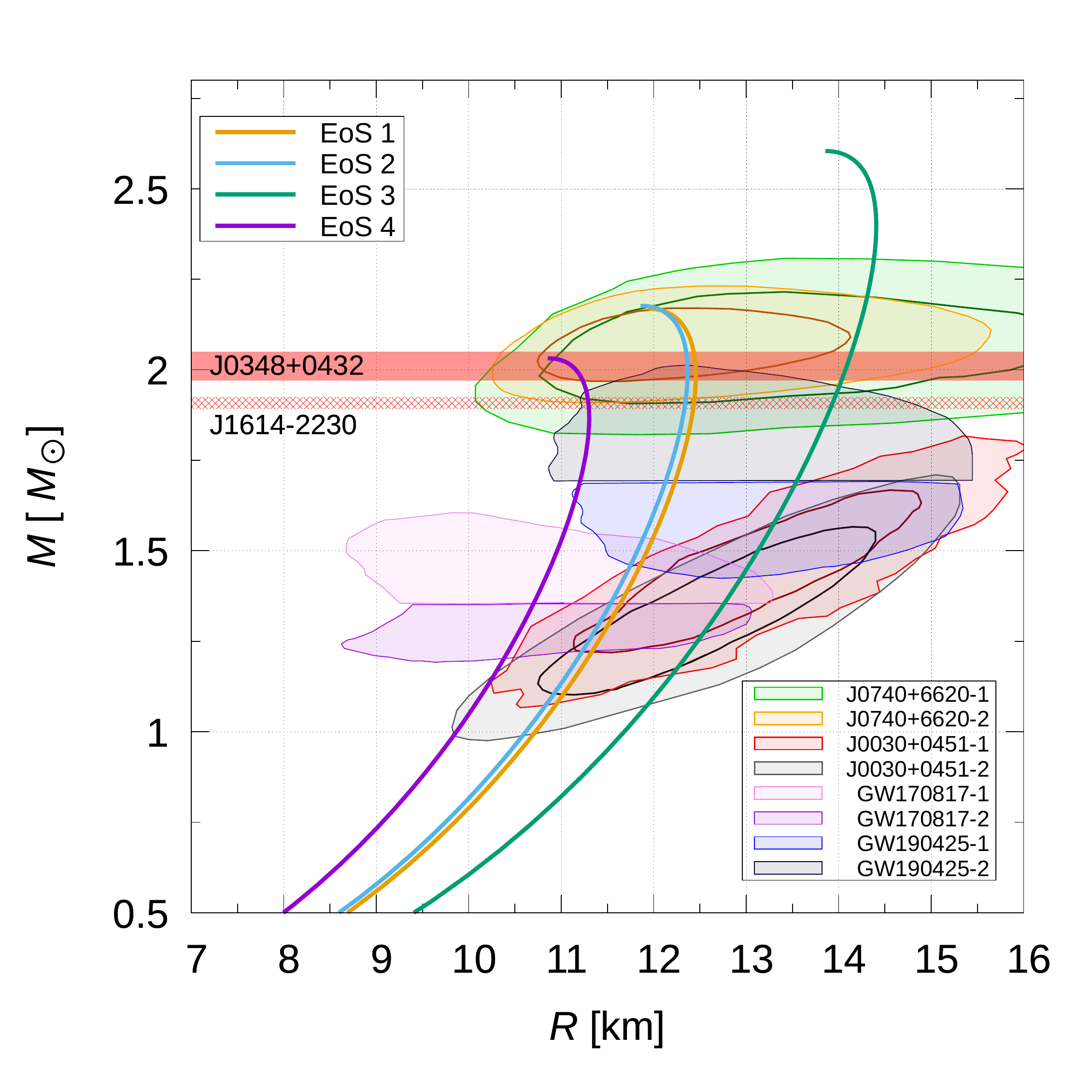}
 \caption{Mass--radius relationships for the sets of Table~\ref{tabla:tablaedes}. Continuous curves represent low-MF QSs, and dotted curves represent magnetars. Due to the insignificant effects of the MF on the QS structure, the two scenarios are indistinguishable (see the enlarged details in Figure~\ref{fig:mr-zoom}). We also present astrophysical constraints from the \mbox{$\sim 2~M_\odot$} pulsars \citep{Arzoumanian:2018tny, Fonseca:2021rfa}, the GW170817 \citep{Abbott:2017oog} and GW190425 \citep{Abbott:2020goo} events, and NICER observations \citep{Miller:2019pjm,Riley:2019anv,Riley:2021anv, Miller:2021tro}.}
 \label{fig:mr_sets}
\end{figure}

Through the selected sets of Table~\ref{tabla:tablaedes}, we studied the structure and oscillation $f$ mode of the QSs considering the two MF parametrizations presented in Table~\ref{table:bcases}: low-MF QSs and magnetars. In all results presented hereafter, the continuous (dashed) curves represent stable solutions for low-MF QSs (magnetars). In Figure~\ref{fig:mr_sets}, we present the mass--radius relationship obtained by integrating TOV equations and the constraints in mass and radius imposed by recent observations: the $\sim 2 M_\odot$ \citep{Arzoumanian:2018tny, Fonseca:2021rfa} pulsars, GW170817 \citep{Abbott:2017oog} and GW190425~\citep{Abbott:2020goo}, and NICER observations \citep{Miller:2019pjm,Riley:2019anv,Riley:2021anv, Miller:2021tro}. It is important to mention that Set~$4$ does not satisfy the restriction imposed by the pulsar J0030+0451. However, we kept Set~$4$ in order to perform a comparative analysis with the other chosen parameter sets.

It can be seen in Figure~\ref{fig:mr_sets} that the Set~$3$ curve reaches a maximum mass of $M_{max}\approx2.6$~M$_\odot$, while the other three EoSs have maximum mass values $M_{max} < 2.2 M_\odot$ (see Table~\ref{tabla:tablaedes}); this result can be explained since, given a fixed $Bag$, an increase in $\Delta$ produces a \emph{stiffer} EoS (and a higher maximum mass); inversely, given a fixed $\Delta$, an increase in $Bag$ produces a \emph{softer} EoS (and a lower maximum mass). This combined effect for Set $3$ (corresponding to the highest $\Delta$ and lowest $Bag$) leads to the highest maximum mass. 

\begin{figure}[H]
%\centering
\includegraphics[width=0.49\linewidth]{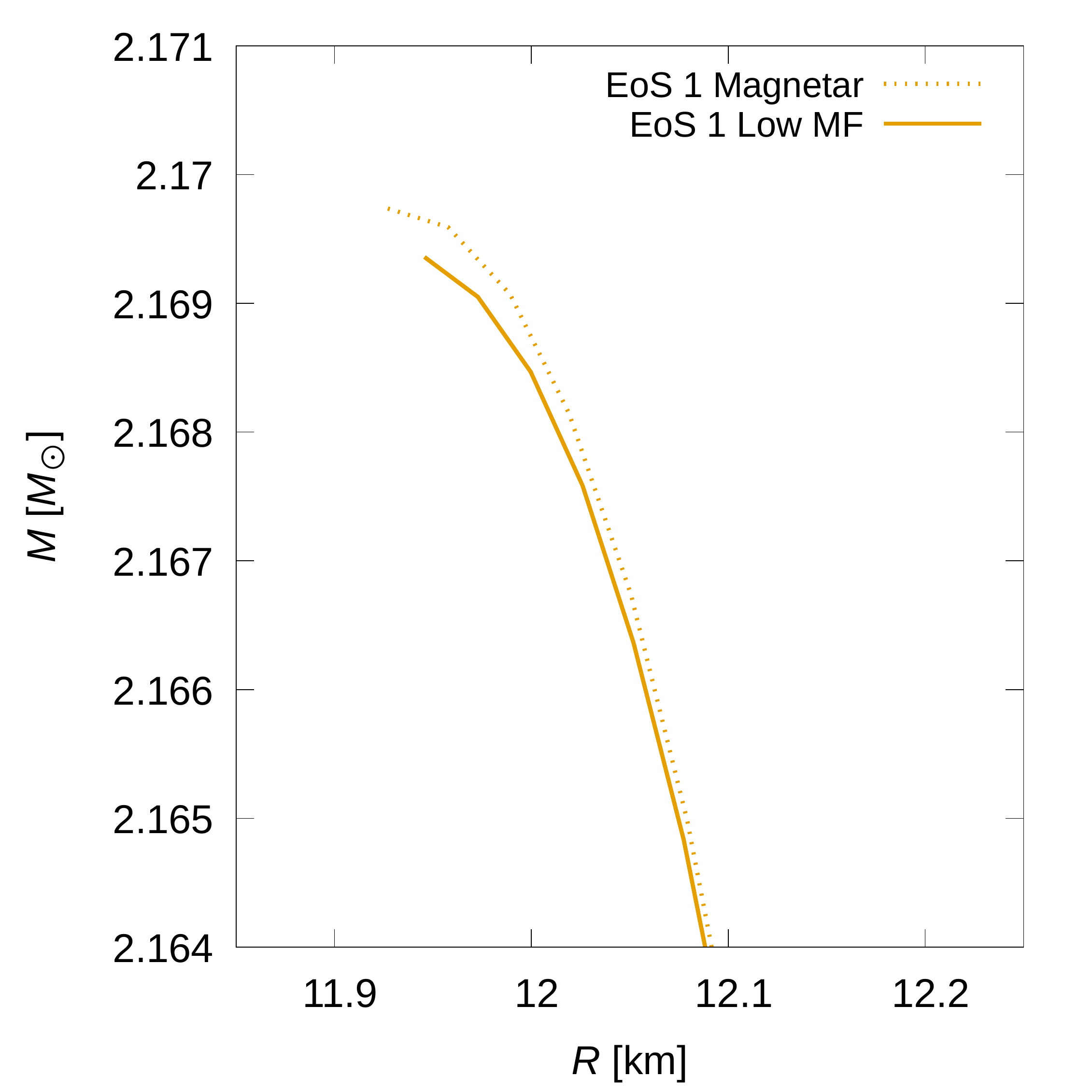} 
\includegraphics[width=0.49\linewidth]{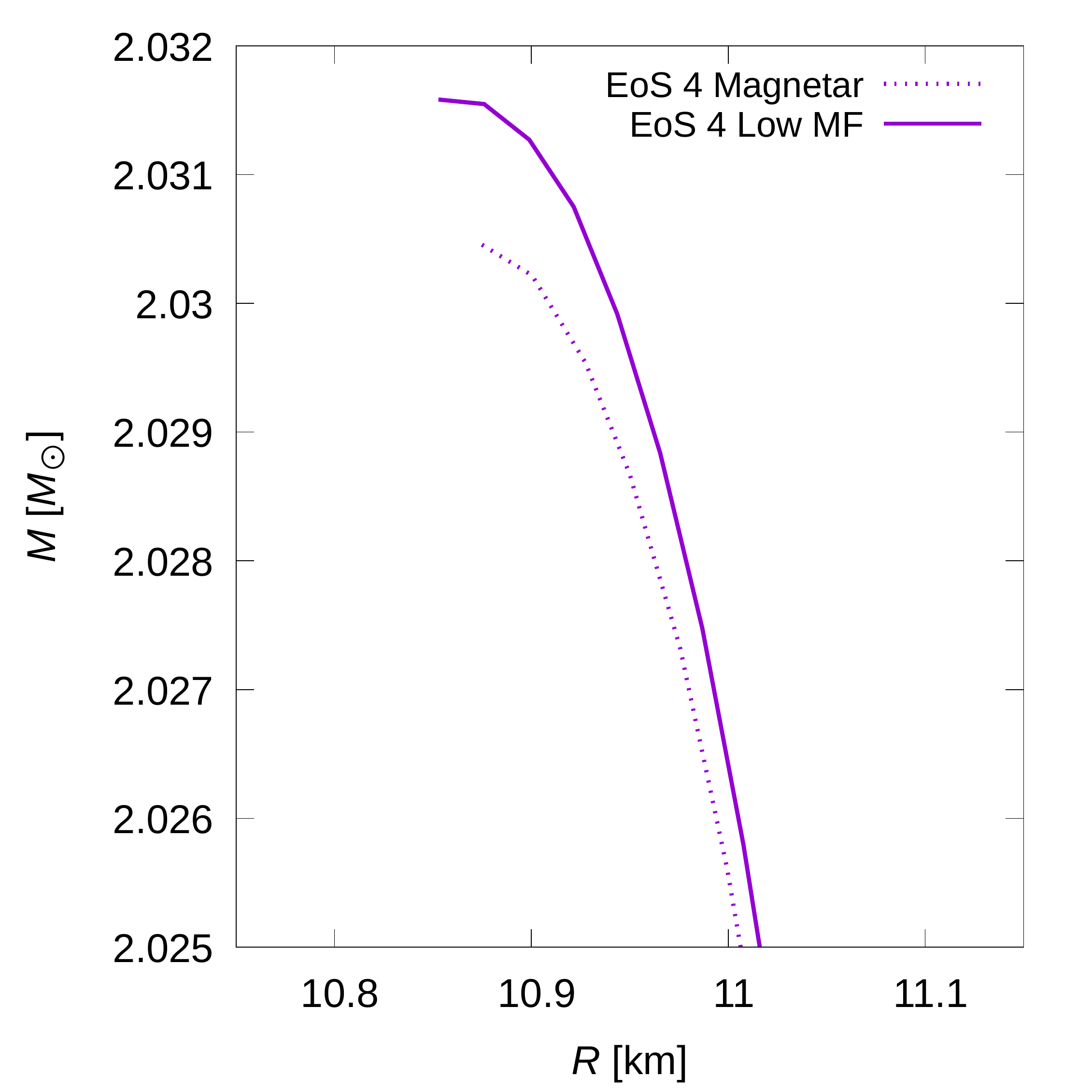} 
\caption{Enlarged zoomed-in view of the mass--radius relationships for the sets of Table~\ref{tabla:tablaedes}. Continuous (dotted) curves represent low-MF (magnetars) QSs. It can be seen that the effect of the MF is negligible. Besides, a higher MF does not imply necessarily a higher value for M$_{max}$. QS families constructed using EoSs~$1$ and $4$ are more sensitive to the variation of the MF strength.}
\label{fig:mr-zoom}
\end{figure}

Furthermore, the effect of the MF is negligible, and this becomes noticeable in the enlarged Figure~\ref{fig:mr-zoom}. This figure shows the detail of continuous and dashed curves around the $M_{max}$ region in the mass--radius plane. We only show the results for Sets~$1$ and $4$, since for Sets~$2$ and $3$, the differences are smaller; in all cases, the differences are less than $1\%$ in the mass at fixed radius. Furthermore, comparing both Sets~$1$ and $4$, it can be seen that a higher MF does not necessarily imply a higher maximum mass.

\vspace{-6pt}

\newpage

On the other hand, to study the oscillation modes of a spherically symmetric body, we must consider not only the oscillations of the star's fluid, but also those that are transferred to space--time through the equations proposed by \citet{Detweiler1985}. The Schwarzschild metric \citep{weber:2017paa} being
\begin{equation}
d s^{2}=-e^{2 \phi(r)} d t^{2}+e^{2 \lambda(r)} d r^{2}+r^{2} d \theta^{2}+r^{2} \operatorname{sen}^{2} \theta d \phi^{2} \, ,
\label{eq:shcw}
\end{equation}
the perturbation functions coupled to the TOV equations are given by \citep{Thorne1967}
%\begin{equation}
%\begin{aligned}
\begin{eqnarray}
H_{1}^{\prime}=&-r^{-1}\left[l+1+2 M e^{\lambda} r^{-1}+4 \pi r^{2} e^{\lambda}(P-\epsilon)\right] H_{1}+r^{-1} e^{\lambda}\left[H_{0}+K\right.\nonumber\\
&-16 \pi(\epsilon+P) V(r)] \, , \nonumber \\
K^{\prime}=& r^{-1} H_{0}+\frac{1}{2} l(l+1) r^{-1} H_{1}-\left[(l+1) r^{-1}-\frac{1}{2} \phi^{\prime}\right] K-8 \pi(\epsilon+P) e^{\lambda / 2} r^{-1} W(r) \,, \nonumber \\
W^{\prime}=&-(l+1) r^{-1} W(r)+r e^{\lambda / 2}\left[\gamma^{-1} P^{-1} e^{-\phi / 2} X-l(l+1) r^{-2} V(r)+\frac{1}{2} H_{0}+K\right] \, , \nonumber \\
X^{\prime}=&-l r^{-1} X+(\epsilon+p) e^{\phi / 2}\left\{\frac{1}{2}\left(r^{-1}-\frac{1}{2} \phi^{\prime}\right) H_{0}+\frac{1}{2}\left[r \omega^{2} e^{-\phi}+\frac{1}{2} l(l+1) r^{-1}\right] H_{1}\right. \nonumber \\
&+\frac{1}{2}\left(2 \phi^{\prime}-r^{-1}\right) K-\frac{1}{2} l(l+1) \phi^{\prime} r^{-2} V-r^{-1}\left[4 \pi(\epsilon+P) e^{\lambda / 2}+\omega^{2} e^{\lambda / 2-\phi}\right.\nonumber \\
&\left.-\frac{1}{2 r^{2}}\left(r^{-2} e^{-\lambda / 2} \phi^{\prime}\right)\right] W \, ,
%\end{aligned}
\label{eq:pertur}
\end{eqnarray}
%\end{equation}
where $H_0$, $H1$, $H2$, and $K$ are time-dependent perturbation functions, $\gamma$ is the adiabatic factor, and $W(r)$ and $V(r)$ are functions characterizing the fluid perturbation \citep{Detweiler1985}. The numerical resolution of Equation~\eqref{eq:pertur} allows us to obtain the oscillation modes of the studied stellar configurations. These modes are known as \emph{quasi-normal modes} (QNMs), since the resulting frequencies are complex, $\omega = 2\pi \nu + i/\tau$, where $\nu$ is the real oscillation frequency and $\tau$ is the oscillation damping time of the corresponding mode. As we already stated, we are interested particularly in the solution of the fundamental $f$ mode. As we are only interested in the pressure $f$ mode, the presence of the magnetic restoration force that induces the magnetic Alfvén modes, or any other magnetic contribution, is not relevant in the perturbation equations. Thus, for the pressure modes, only the MF effects on the pressure are relevant, and we introduce them trough the magnetized EoS.

\begin{figure}[H]
% \centering
 \includegraphics[width=0.49\linewidth]{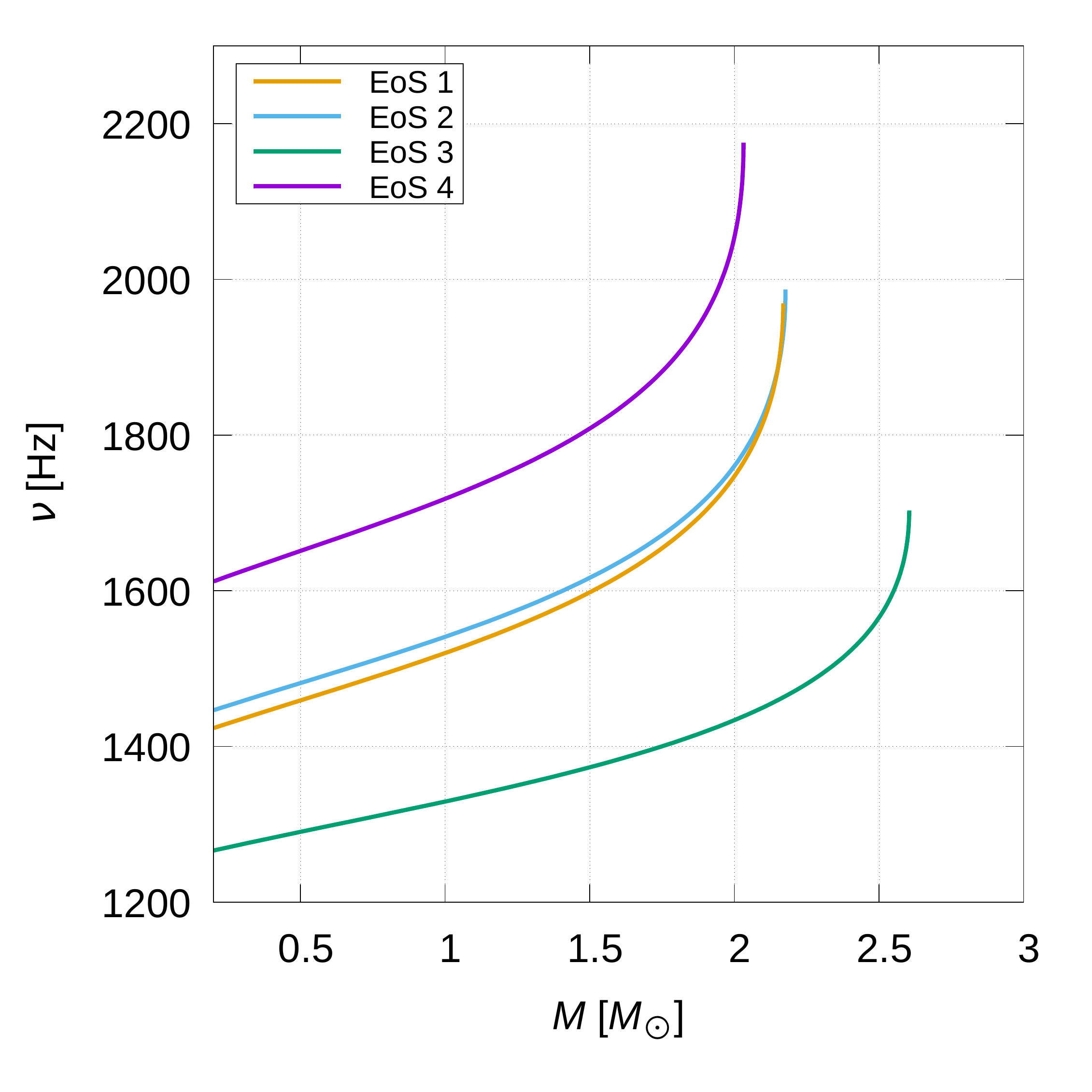}
 \includegraphics[width=0.49\linewidth]{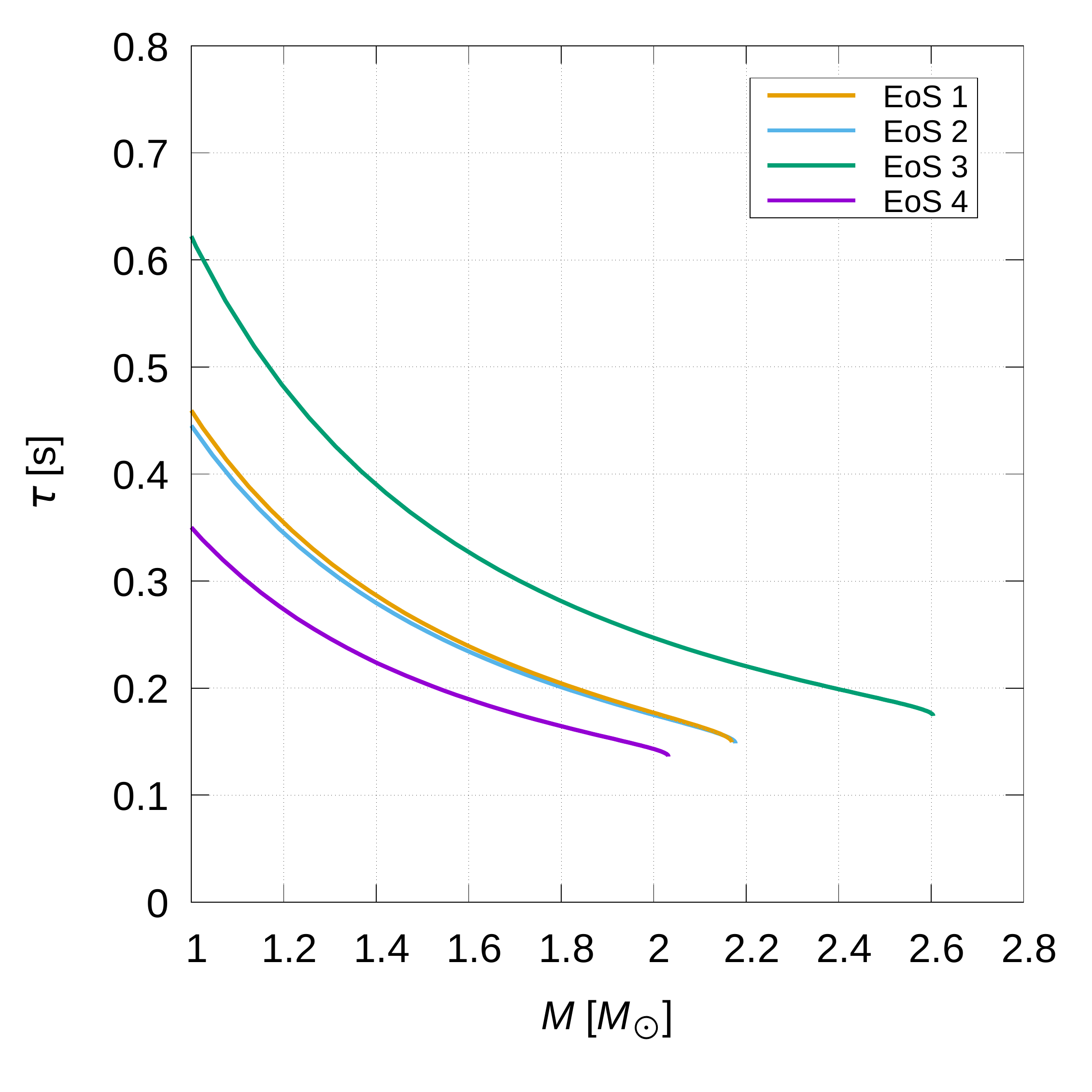}
 \caption{$\nu$--mass (left panel) and $\tau$--mass (right panel) relationships for the sets of Table~\ref{tabla:tablaedes}. Continuous curves represent low-MF QSs, and dotted curves represent magnetars. Due to the negligible effects of the MF on the QS structure, both astrophysical scenarios are indistinguishable (see enlarged Figure~\ref{fig:frec-zoom} for details). It can be seen that the EoS producing the highest maximum mass results in the lowest frequency values and longest damping times.}
 \label{fig:fmode}
\end{figure}

\begin{figure}[H]
%\centering
\includegraphics[width=0.49\textwidth]{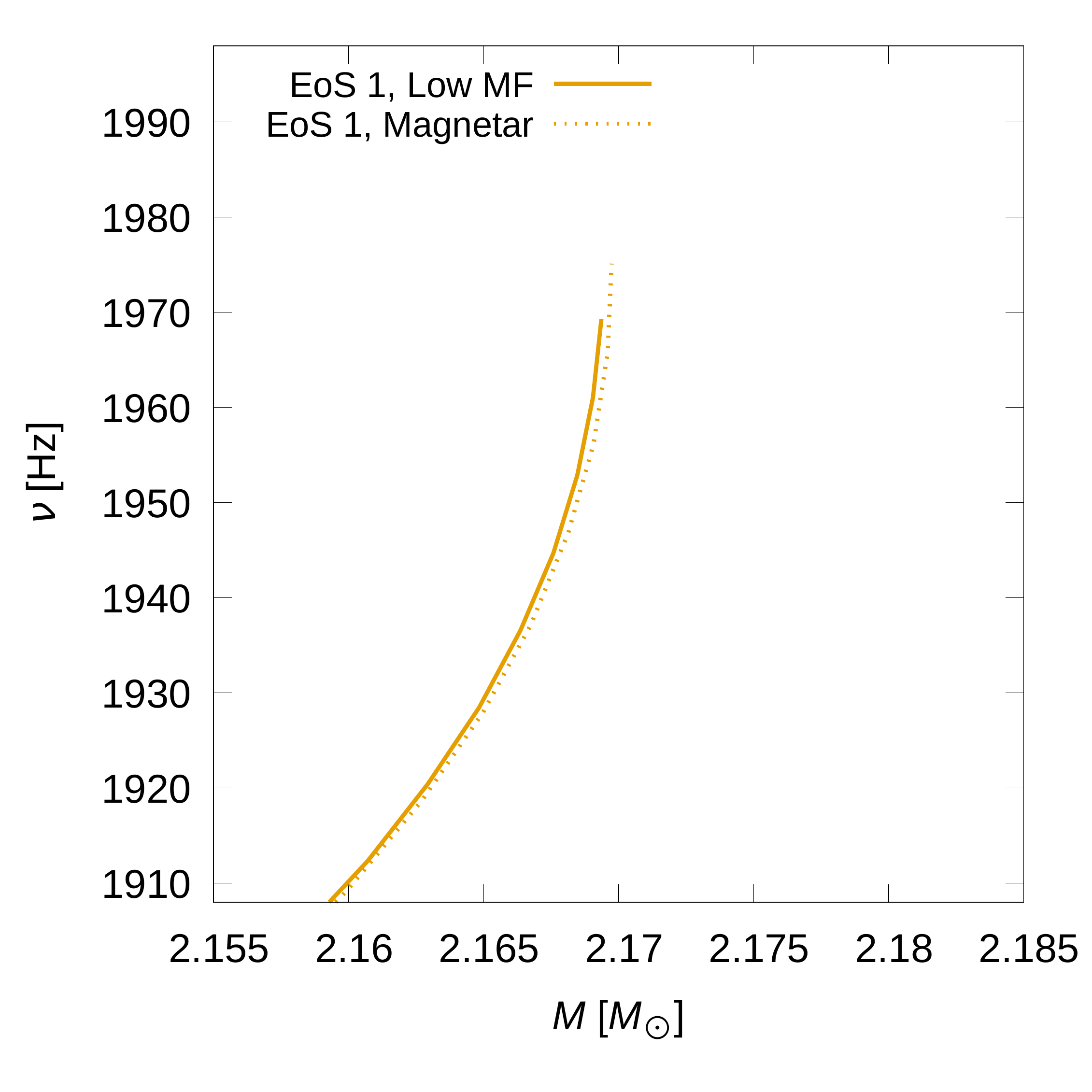} 
\includegraphics[width=0.49\textwidth]{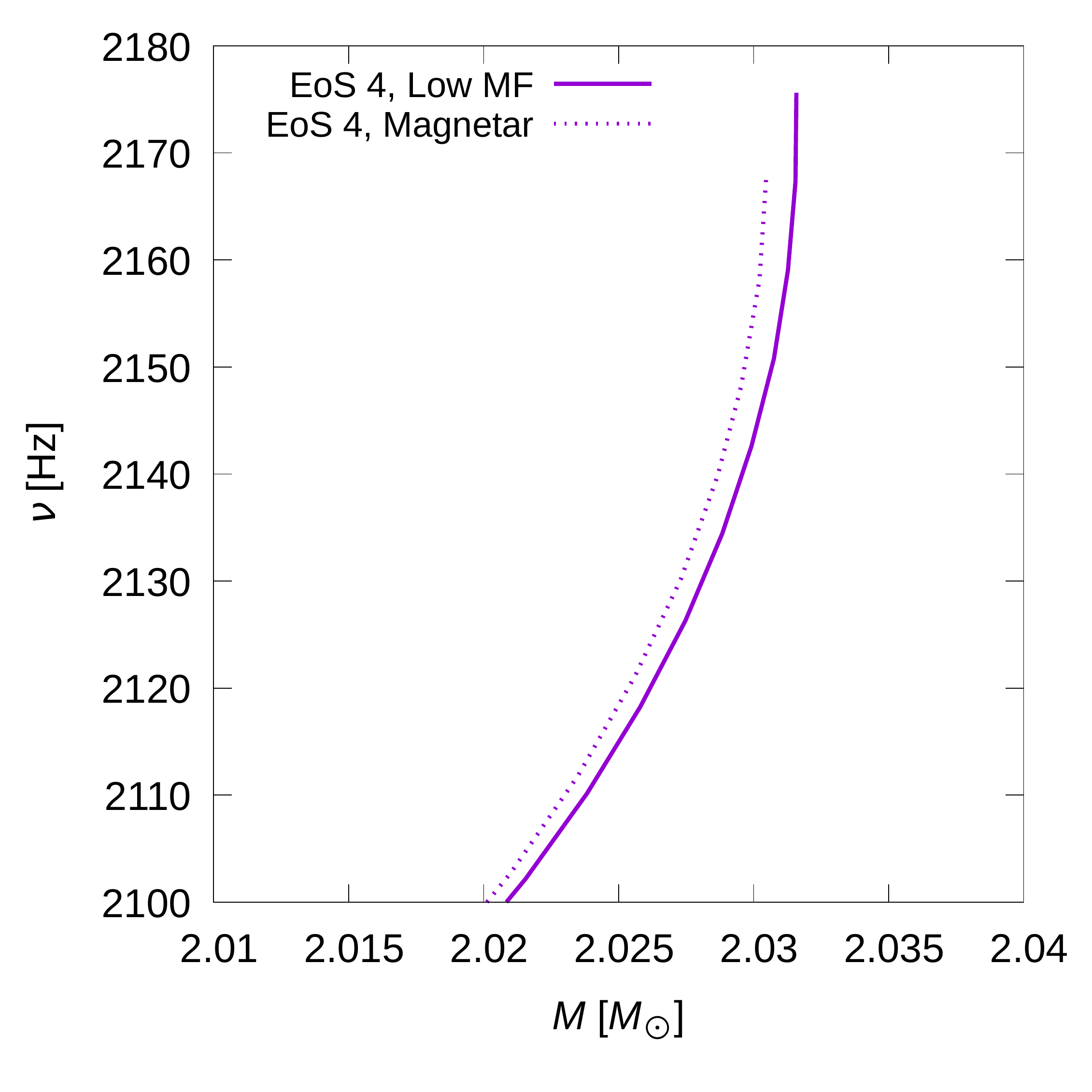} 
\label{fig:frec-cuarta}
\caption{Enlarged zoomed-in view of the frequency--mass relationships for the sets of Table~\ref{tabla:tablaedes}. Continuous (dotted) curves represent low-MF (magnetars) QSs. The effect of the MF is negligible; as we already pointed out in the mass--radius relationship, the QS families constructed using EoSs~$1$ and $4$ are more sensitive to the variation of the MF strength.}
\label{fig:frec-zoom}
\end{figure}

In Figure~\ref{fig:fmode}, we present the results for the $f$ mode in the $\nu$-mass (left panel) and $\tau$-mass (right panel) planes. In the $\nu$-mass plane, it can be seen that oscillation frequencies are not only increasing with the mass of the stellar configurations, but also that families of QSs with higher maximum masses have lower oscillation frequencies; e.g., for a QS of $2 M_\odot$, $\nu$~$\sim 2100$~Hz for Set~$4$, $M_{max}\approx2.03 M_\odot$ and $M_{max}\approx2.17M_\odot$ correspond to $\nu$ $\sim 1780$~Hz for Sets~$1$ and $2$, and $M_{max}\approx2.6 M_\odot$ corresponds to $\nu$ $\sim 1450$~Hz for Set~$3$. Comparing Sets~$1$ and $3$, it follows that an increase in $\Delta$ generates lower frequencies; comparing Sets~$3$ and $4$, we observe that an increase in the $Bag$ produces an increase in the frequency values. In particular, we also calculated the $f$ mode for the non-superconducting magnetized strange quark matter ($\Delta=\mu_3=\mu_8=0$) with $Bag = 45$~MeV/fm$^3$. This particular choice allows us to compare the non-superconducting case with Sets~$1$ and $3$. As we have already shown, an increase in $\Delta$ implies lower $f$ mode frequencies, and this behavior is also valid in the limit of $\Delta = 0$. For a given $Bag$ value, the non-superconducting case has the highest frequency, and the appearance of the superconducting phase implies a decrease in the frequency. To quantify this comparison, we calculated the percentage change for the $M_{max}$ QS as an indicator of the whole family's results; if we compare the non-superconducting case with those of the $\Delta=10$~MeV and $\Delta=90$~MeV results, $\nu$ has a shift of $\sim 1\%$ and $\sim 15\%$, respectively. On the other hand, analogous to what happens in the mass--radius plane, the effect of an intense MF is practically negligible over the values of the $f$ mode oscillation frequencies (see the enlarged Figure~\ref{fig:frec-zoom} for details of Sets $1$ and $4$).

Considering all the sets, for masses $1.0$--$2.6 M_{\odot}$, we obtained frequencies in the range $1200$--$2200$~Hz. Previous works obtained qualitatively similar results for purely hadronic, hybrid, or quark stars \citep{Sotani:2003noo,Benhar:2007qmi,Flores:2017ccf,Tonetto:2020dgm,Rodriguez:2021hsw,Flores:2020gws}; thus, it is not possible to distinguish among hadronic, hybrid, or quark EoSs through the eventual detection of the $f$ mode frequency. In particular, \citet{Flores:2017ccf} studied color superconducting QSs at zero M,F and their results are in agreement with those obtained in this work for the frequencies associated with the $f$ mode.

\vspace{-6pt}

For the $\tau$--mass relationship, presented in the right panel of Figure~\ref{fig:fmode}, our results are in the range of $0.1$-$0.6$~seconds, also in agreement with the previously mentioned works \citep{Sotani:2003noo, Benhar:2007qmi,Flores:2017ccf,Tonetto:2020dgm,Flores:2020gws}. Contrary to the $\nu$ behavior, $\tau$ decreases with the mass, and the curves of the QS families with a higher maximum mass correspond to higher $\tau$. Besides, an increase in $\Delta$ produces increased%please ensure that the original meaning is retained
 $\tau$, while an increase in $Bag$ causes a reduction in $\tau$. In the non-superconducting case, $\tau$ has the lowest value, and it increases when we consider color superconductivity; comparing the non-superconducting result with Sets~$1$ and $3$ for $M_{max}$, $\tau$ has a less than $1\%$ shift between the $\Delta = 0$ and $\Delta = 10$~MeV cases and a $\sim 17\%$ shift between the $\Delta = 0$ and $\Delta = 90$~MeV cases. The effect of the MF strength is negligible.

\begin{figure}[H]
% \centering
 \includegraphics[width=0.49\linewidth]{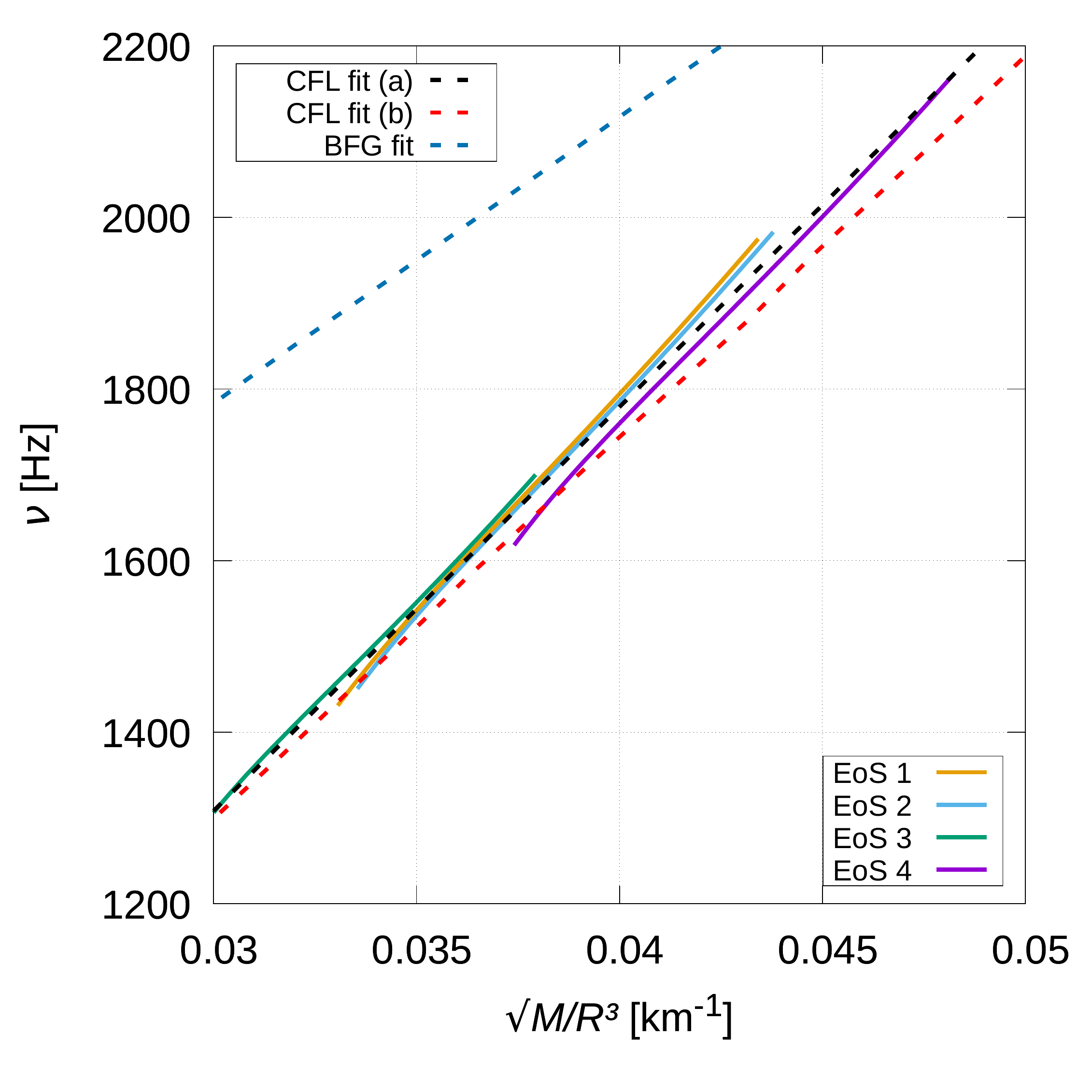}
 \includegraphics[width=0.49\linewidth]{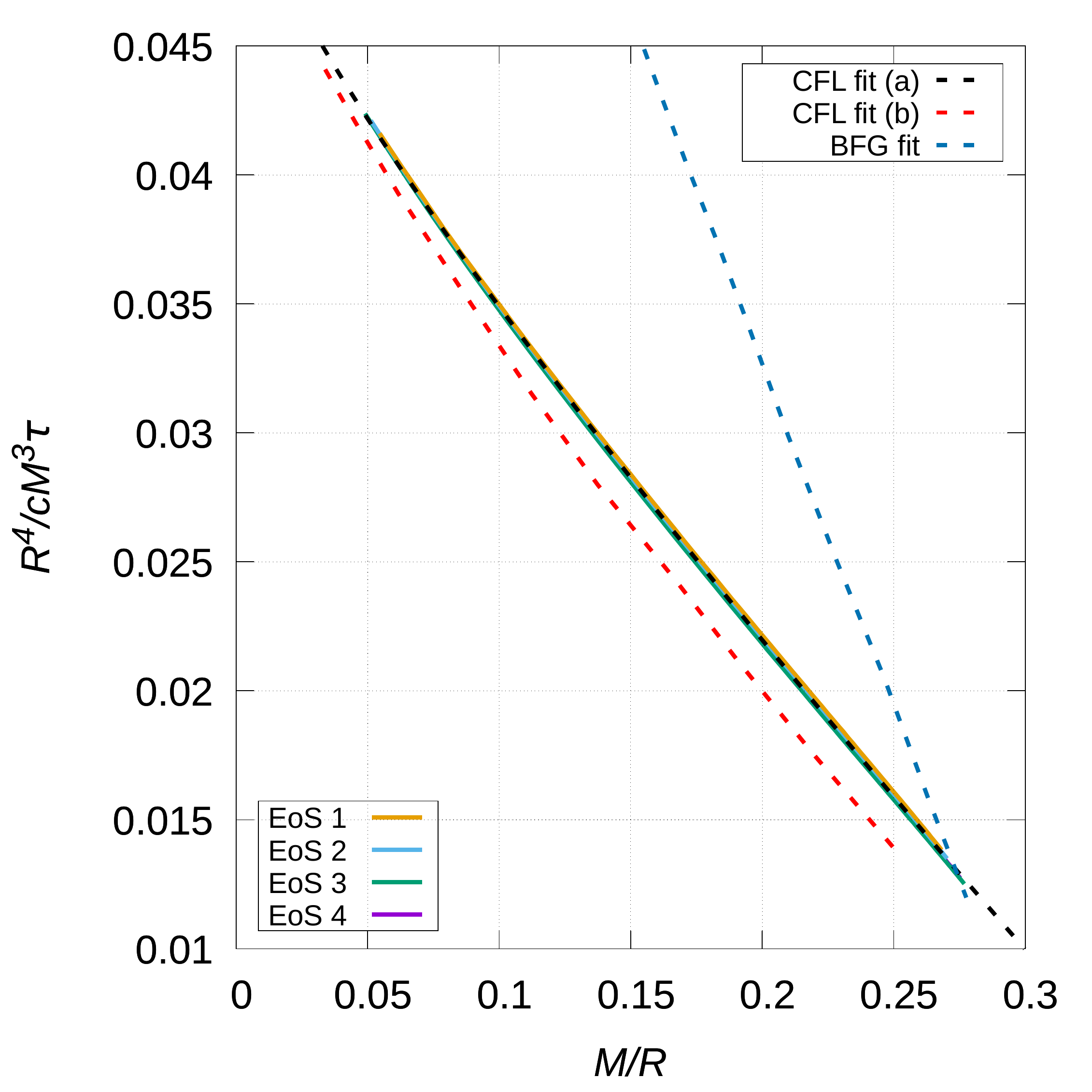} 
 \caption{Universal relationships proposed by \citet{Andersson:1998tgw} for the $f$ mode, for $\nu$ as a function of the mean density (left panel) and for $\tau$ as a function of the compactness (right panel). In both panels, we also show the BFG \citep{Benhar2004} and two CFL fits; CFL fit (a) corresponds to the fit calculated in this paper and CFL fit (b) to Flores and Lugones' fit \citep{Flores:2017ccf}. Our fit lies, predictably, close to the CFL fit (b).}
 \label{fig:univ}
\end{figure}

On the other hand, there exist \emph{universal relationships} associated with the oscillation frequencies and damping times for the $f$ mode \citep{Andersson:1998tgw}. In particular, there are empirical relations relating the frequency and damping time to the mass and radius of a stellar object given by
\begin{equation}\label{eq:relf}
 \nu = a_1 + b_1\sqrt{\frac{\textrm{M}}{\textrm{R}^3}} \, ,
\end{equation}
\begin{equation}\label{eq:reltau}
 \frac{\textrm{R}^4}{\textrm{M}^3\tau} = a_2 + b_2\sqrt{\frac{\textrm{M}}{\textrm{R}}} + c_2 \frac{\textrm{M}}{\textrm{R}} \, .
\end{equation}
In order to analyze how the magnetized color superconducting EoS fits the universal relationships, we used two fits in Equations~\eqref{eq:relf} and \eqref{eq:reltau}: BFG fit for hadronic matter \citep{Benhar2004} and CFL fit for quark matter \citep{Flores:2017ccf} (see Table~\ref{tabla:fits} for details). If the universality of these relationships holds, the detection of both the frequency and damping time of the fundamental mode of a given compact object allows us to infer the properties of the star such as mass and radius, independently of the EoS used to describe its composition. In Figure~\ref{fig:univ}, we present our results and the mentioned fits for the $\nu$ (left panel) and $\tau$ (right panel) universal relationships. As can be seen, the results for all the sets are grouped in very narrow regions along the CFL fit. The main difference between the EoSs used in this work and the EoSs of \cite{Flores:2017ccf} is the strange quark mass, $m_s$. We added a new fit (detailed in the last column of Table~\ref{tabla:fits}) corresponding to $m_s$ = 96 MeV, used in Equation~\eqref{eq:relf}, which matches very well with all the EoSs of the present work. \citet{Flores:2017ccf} used \mbox{$m_s$ = 150 MeV}. Moreover, they considered zero magnetic field and massless $u$ and $d$ quarks, and they did not take into account the color chemical potentials, $\mu_3$ and $\mu_8$. The difference in the coefficient values between both fits are not significant, indicating the usual dispersion from the universal relationships for different EoS models. Therefore, within our model, the universal relations developed by \citet{Andersson:1998tgw} are valid and the results obtained are in agreement with the fit for QSs presented by \citet{Flores:2017ccf}.

\begin{table}[H]
\centering
%\resizebox{0.6\textwidth}{!}
\caption{Parameters for the BFG \citep{Benhar2004} and CFL \citep{Flores:2017ccf} fits related to the $f$ mode in Equations~\eqref{eq:relf} and~\eqref{eq:reltau}. In the third column, we show the parameters fitting our magnetized CFL results.}
\label{tabla:fits}
{
\begin{tabular}{cccccc}
\toprule
fit & \begin{tabular}[c]{@{}c@{}} $a_1$ {(}Hz{)}\end{tabular} & \begin{tabular}[c]{@{}c@{}} $b_1$ $($km Hz$)$ \end{tabular} & \begin{tabular}[c]{@{}c@{}} $a_2$ \\ \end{tabular} & \begin{tabular}[c]{@{}c@{}} $b_2$ \\ \end{tabular} & \begin{tabular}[c]{@{}c@{}} $c_2$ \\ \end{tabular}\\ \midrule
BFG & $790$ & $33 \times 10^3$ & $8.7 \times 10^{-2}$ & 0 & $-0.271$ \\ 
CFL \citep{Flores:2017ccf} & $-23$ & $44.11 \times 10^3$ & $0.0553$ & $-0.0466$ & $-0.0725$ \\ 
CFL (this paper) & $-103$ & $47.07 \times 10^3$ & $0.0534$ & $-0.0302$ & $-0.0897$ \\
\bottomrule
\end{tabular}
}

\end{table}

\section{Summary and Discussion}\label{sec:conclus}

Using the MIT bag model, including MF and CFL color superconductivity effects, we modeled magnetized color superconducting quark stars and calculated their oscillation $f$ mode and associated frequencies and damping times. For the treatment of the MF, we adopted the chaotic approximation and a functional profile with realistic surface and central MF values for the two astrophysical scenarios considered: low-MF and magnetar QS. We constructed the stability window for magnetized superconducting quark matter and took into account the constraint of massive pulsars to select four sets of parameters that represent the qualitative behavior of the model used. For these four EoSs, we analyzed the mass--radius diagram considering the last constraints on neutron stars given by the GW events $170817$ and $190425$ and NICER observations. In addition, we verified the universal relations for the frequencies and damping times associated with the oscillation $f$ mode for the chosen EoSs.
 
The results show that the inclusion of a superconducting color term related to di-quark formation in the EoS produces significant effects, not only on the stability window of strange quark matter, but also on astrophysical quantities, such as the maximum mass, radii, or oscillation $f$ mode of QSs.

On the other hand, the differences of the EoSs in the cases of low and intense MFs are practically negligible. This leads to the fact that the results of mass, radius, and oscillation frequencies obtained do not present differences when comparing both astrophysical scenarios. In particular, for the $f$ mode, our results are in agreement with the work by \citet{Lander:2010oor}, where, under a Newtonian approach for NSs, they predicted a small shift, proportional to the influence of the pure magnetic pressure, $\propto B^2$, on the total pressure. It is worth mentioning that, if we had used a realistic MF internal distribution, such as the polynomial profiles by \citet{Dexheimer:2017wis, Chatterjee:2019mfd} or the ones arising from MHD simulations \citep{Pili:2014aem}, together with the magnetar observed surface value $B \sim 10^{15}$~Gauss, these results would not have changed qualitatively. The fact that the considered quark EoSs only show significant shifts for $B \gtrsim 10^{19}$~Gauss results in that all the discussed parametrizations would have shown negligible MF effects.

Except in the case of EoS~$4$, the families of stars built with EoSs~$1$, $2$, and $3$ satisfy the observational restrictions imposed by GWs, NICER, and massive pulsars. However, EoS~$4$ is useful to identify the behavior of the \textit{Bag} and $\Delta$ parameters compared to the other considered EoSs.

The frequencies and damping times of the QNM $f$ mode are in agreement with the results of purely hadronic, hybrid, and quark stars from previous works \citep{Sotani:2003noo, Benhar:2007qmi,Flores:2017ccf,Tonetto:2020dgm,Flores:2020gws}. This implies that if the emission of the $f$ mode could be detected, it would be impossible to distinguish whether these signals come from a purely hadronic, hybrid, or quark star by simply determining the frequency or damping time. We also observed that \textit{stiffened} EoSs, those that provide mass--radius curves with higher maximum masses, give rise to lower oscillation frequencies and higher damping times when compared with \textit{softened} EoSs.

The QNMs obtained fit very well with the universal relationships corresponding to QSs' EoSs. The analysis of these empirical relations would allow not only obtaining structural parameters of the detected object independent of the EoS, such as mass or radius, but also classifying such a compact object and finding possible observational evidence of quark matter in its composition.

It is expected that the third-generation GW observatories, such as the Einstein Telescope, could detect $f$ mode emissions of isolated NSs; we hope that our results can be tested once these detectors start operating in the next few years.

%%%%%%%%%%%%%%%%%%%%%%%%%%%%%%%%%%%%%%%%%%
\vspace{6pt} 

%%%%%%%%%%%%%%%%%%%%%%%%%%%%%%%%%%%%%%%%%%
%% optional
%\supplementary{The following supporting information can be downloaded at: \linksupplementary{s1}, Figure S1: title; Table S1: title; Video S1: title.}


%%%%%%%%%%%%%%%%%%%%%%%%%%%%%%%%%%%%%%%%%%
\newpage
\section*{Acknowledgements}
The authors contributed equally to the theoretical and numerical aspects of the work presented in this paper. All authors have read and agreed to the published version of the~manuscript.

This research was supported by CONICET and UNLP, Grant Numbers PIP 0714 and G 157.
The authors thank the anonymous Referees for their corrections and suggestions, which helped improve the quality of the manuscript. M.O.C. and M.M. are fellows of Consejo Nacional de Investigaciones Cient{\'i}ficas y T{\'e}cnicas (CONICET). M.M. and M.G.O. thank CONICET and the Universidad Nacional de La Plata (UNLP) for financial support. L.T. thanks the Italian Istituto Nazionale di Fisica Nucleare (INFN) under Grant TEONGRAV.%}

%\conflictsofinterest{
The authors declare no conflict of interest.%} 

\renewcommand{\bibname}{References}
\bibliographystyle{apalike} % estilo de la bibliografía.
\bibliography{Bibliography.bib}

\end{document}